\documentclass[12pt]{iopart}
\usepackage{iopams}
\usepackage{graphicx}
\usepackage{color}
\usepackage{bm}

\begin{document}

\title{Von Neumann spin measurements with Rashba fields}
\author{E. Ya. Sherman$^{a,b}$ and D. Sokolovski$^{a,b}$}
%\date{\today}
\address{
$^a$ Departmento de Qu\'imica-F\'isica, Universidad del Pa\'is
Vasco, UPV/EHU, E-48080 Leioa, Spain\\
$^b$ IKERBASQUE Basque Foundation for Science, Bilbao, Spain}

\ead{evgeny.sherman@ehu.es}

\begin{abstract}
We show that dynamics in spin-orbit coupling field simulates
the von Neumann measurement of a particle spin. We
demonstrate how the measurement influences the spin and coordinate evolution
of a particle by comparing two 
examples of such a procedure. First example is a simultaneous measurement of spin
components, $\sigma _{x}$ and $\sigma _{y}$, corresponding to non-commuting
operators, which cannot be accurately obtained together at a given time instant
due to the Heisenberg uncertainty ratio.
By mapping spin dynamics onto a spatial walk such a procedure 
determines measurement-time averages of $\sigma _{x}$ and $\sigma _{y}$,
which already can be precisely evaluated in a single short-time measurement. The other, 
qualitatively different, example is the spin of a one-dimensional particle in a magnetic field. 
Here the outcome depends on the angle between
the spin-orbit coupling and magnetic fields.  These results can be applied to 
studies of spin-orbit coupled cold atoms and electrons in solids.
\end{abstract}

\maketitle

\newpage 

\section{Introduction: The spin-orbit coupling Hamiltonian and von Neumann measurement.}

Unusual properties \cite{Winker03,Zutic04,Dyakonov08,Wu10,GlazovReview}, simplicity,  
and ability to manipulate the strength \cite{Karimov}
put the spin-orbit coupling (SOC) in the focus of many research fields, where the mutual dependence
of spin and coordinate motion is important \cite{Rokhinson}.
For several decades, solids and solid-state structures were the systems to
study the effects of spin-orbit coupling for electrons  \cite{BychkovRashba} and holes \cite{Winker}.
Recently, at least two completely new classes of systems with spin-orbit coupling were discovered
and became new research fields.
The first class is cold atoms in highly coherent laser fields \cite{Wang10,Stanescu,Lin,Wang,Sommer,Cheuk,Liu}. 
For both types of particles, in addition to the SOC, an effective magnetic field can be
produced optically. Studies of spin-orbit coupled cold atoms are related 
to the properties of new phases (see e.g. \cite{Ozawa,Xiong,ZhaiReview}) and 
macroscopic spin dynamics \cite{Tokatly,Natu,Wu13}. 
The other class is topological insulators \cite{Hasan}, 
considered as promising elements for the spintronics applications.   
 
The very general character of the spin-orbit coupling should have consequences 
for the {fundamental} quantum mechanics of single particles and their ensembles
and {stimulate the search for these consequences in experimentally realizable system.} 
{Motivated by the interest in the observability of the effects of spin-orbit 
coupling on the basic quantum mechanical level, here we show that and how it is 
directly related to the quantum measurement of spin 1/2 in terms of the procedure proposed by
von Neumann.} We begin with a general Hamiltonian (assume $\hbar\equiv1$) 
of spin-orbit coupling linear in the two-dimensional particle momentum: 
\begin{eqnarray}
\hat{H} &=&\hat{H}_{0}+\hat{H}_{\rm so}+\hat{H}_{Z},  \label{Hamiltonian} \\
\hat{H}_{0} &=&\frac{k_{x}^{2}+k_{y}^{2}}{2M},  \label{Hamiltonian0} \\
\hat{H}_{\rm so} &=&\alpha(t)\left[ \nu _{x}k_{x}
\left(\mathbf{h}_{{\mathbf{x}}}\cdot{\bm\sigma }\right) +
\nu _{y}k_{y}\left( \mathbf{h}_{{\mathbf{y}}}\cdot{\bm\sigma }\right) \right] ,
\qquad \hat{H}_{Z}=\frac{\Delta }{2}\left( \mathbf{b}\cdot{\bm\sigma }\right),
\label{HamiltonianSO}
\end{eqnarray}
where $\hat{k}_{x}=-i\partial /\partial x - {\cal A}_{x}$ and $\hat{k}_{y}=-i\partial/\partial y - {\cal A}_{y}$ 
with corresponding components of gauge potential ${\cal A}_{x,y},$ $M$ is the effective mass,
and ${\bm\sigma}$ is the Pauli matrices pseudovector.
Here $\alpha(t)$ is spin-orbit coupling
parameter, in general time-dependent, ${\bm\nu}$ is a two-dimensional vector of unit length, 
$\mathbf{h}_{{\mathbf{x}}}$ and $\mathbf{h}_{{\mathbf{y}}}$ are the unit vectors corresponding to the type of 
spin-orbit coupling, $\Delta$ is the Zeeman splitting, and 
$\mathbf{b}$ is the unit vector of the direction of magnetic
field. The conventional Rashba coupling is given by 
${\bm\nu} =\left(1,1\right)/\sqrt{2},$ $\mathbf{h}_{{\mathbf{x}}}=\mathbf{y},$ $\mathbf{h}_{{\mathbf{y}}}=-\mathbf{x}$, while
the Dresselhaus couping is given by
${\bm\nu} =\left(1,1\right)/\sqrt{2},$ $\mathbf{h}_{{\mathbf{x}}}=\mathbf{x},$ and $\mathbf{h}_{{\mathbf{y}}}=-\mathbf{y}$
($\mathbf{x}$ and $\mathbf{y}$ are the corresponding coordinate system vectors).
The value of $\alpha$ is in the range from 1-10 cm/s for cold 
atoms to 10$^{6}$ cm/s for electrons in semiconductor nanostructures and 
10$^{7}$-10$^{8}$ cm/s for and topological insulators, where the kinetic energy
vanishes \cite{Hasan}, and surfaces with extremely strong spin-orbit coupling \cite{Gierz,Yaji,Mathias,Ibanez}. 

As a result, even at $\Delta=0$, spin rotates around the axis which direction depends on $\mathbf{k}.$ 
%A special case is $[\mathbf{h}_{\left[ x\right] },%
%\mathbf{h}_{\left[ y\right] }]=0$, where the spin-orbit field can be removed
%by a gauge transformation \cite{Tokatly10}, spin precession axis becomes ${\bf k}-$ independent, and 
%spin helix state can be formed \cite{Koralek09}.
The operator of particle's velocity depends on the orientation of the particle's spin, 
\begin{equation}
\hat{v}_{x}=i\left[ H,x\right] =\frac{k_{x}}{M}+\alpha(t) \nu _{x}\left( 
\mathbf{h}_{\mathbf{x}}\cdot{\bm\sigma }\right) ,
\quad \hat{v}_{y}=i\left[H,y\right] =\frac{k_{y}}{M}+\alpha(t)\nu _{y}\left(\mathbf{h}_{\mathbf{y}}\cdot{\bm\sigma}\right).  
\label{velocities}
\end{equation}
The velocity components do not commute with each other if the
cross product of $\mathbf{h}_{\mathbf{x}}$ and 
$\mathbf{h}_{\mathbf{y}}$ is not zero. Even if it is zero, but 
$\mathbf{h}_{\mathbf{x}},\mathbf{h}_{\mathbf{y}},$ and $\mathbf{b}$ 
are not collinear, the velocity components do not
commute with the total Hamiltonian, complicating strongly the orbital dynamics.  

Now we can see a connection between Hamiltonian (\ref{Hamiltonian}) and a quantum spin measurement.
Consider a quantum system characterized 
by a multicomponent operator $\left( \hat{O}%
_{x},\hat{O}_{y},\hat{O}_{z}\right)$ coupled to momentum  
$\hat{\mathbf{P}}$ of another system, ``pointer'', as $\sum\kappa_{ji}\hat{P}_{j}\hat{O}_{i},$
where $\kappa_{ji}$ is the corresponding coupling strength.   
This coupling is presented as $\left(-i\partial/\partial X_{j}\right)\hat{O}_{i},$ 
where $\hat{\mathbf{X}}$ is the pointer position,
and the Hamiltonian for the free pointer is $H_{\rm pnt}(\hat{\mathbf{P}},\hat{\mathbf{X}}).$ 
This coupling causes evolution of the quantities, characterized by
operator $\hat{\mathbf{O}}$ and, at the same time, makes the
corresponding dynamics visible by mapping it on the pointer position in the
coordinate space. Thus, by tracing the pointer position, one can expect tracing
the motion of the operator components of $\hat{O}_{i}.$ In the case of
spin-orbit coupling, $\hat{\mathbf{P}}$ is the particle momentum, and $%
\hat{\mathbf{O}}$ represents the spin components,
as can be seen from Eqs.(\ref{Hamiltonian})-(\ref{HamiltonianSO}). This simple
observation, being the idea behind the von Neumann measurement, can have
interesting consequences, {including entanglement of the spin and coordinate degrees 
of freedom and, correspondingly, a spin dephasing in the measurement procedure, 
for experimentally realizable systems. We will describe these effects in this paper.} Moreover, we will 
demonstrate, that the dynamics of the pointer due to the Hamiltonian $%
H_{\rm pnt}(\hat{\mathbf{P}},\hat{\mathbf{X}}),$ e.g., its kinetic energy, influences
the measurement procedure and its accuracy. This approach corresponds to the
measurements by solving dynamical models, with some of them 
recently reviewed in Ref. \cite{Allahverdyan13}.

To perform a spin measurement, we choose the spin-orbit coupling $\alpha (t)$ switched on 
for a finite time interval 
\begin{equation}
\alpha (t)=\alpha \times \left\{ 
\begin{array}{ll}
1, & 0\le t\le T \\
0, & t<0,\quad t>T,
\end{array}
\right.
\label{g_of_t}
\end{equation}
making the Hamiltonian time-independent during the measurement. 
As a result, in the measurement procedure, evolution of 
the initial state ${\bm\Psi}(\mathbf{r}|0),$
where $\mathbf{r}$ is the position, and ${\bm\Psi}(\mathbf{r}|t)$ 
is the two-component spinor wavefunction is given by: 
\begin{equation}
{\bm\Psi}(\mathbf{r}|t)=\exp(-i\hat{H}t){\bm\Psi}(\mathbf{r}|0).
\label{expHt}
\end{equation}
For a  translational invariant system of our interest, 
it is convenient to represent Eq.(\ref{expHt}) in the form:
\begin{equation}
\hspace{-2cm}
{\bm\Psi}(\mathbf{r}|t)=
\int\mathbf{G}\left(\mathbf{r}-\mathbf{r}^{\prime}|t\right) 
{\bm\Psi }(\mathbf{r}^{\prime }|0)d^{D}r^{\prime }
=\int \mathbf{G}\left(\mathbf{k}|t\right){\bm\Psi }(\mathbf{k}|0)
e^{i\left(\mathbf{k}\cdot\mathbf{r}\right)}\frac{d^{D}k}{(2\pi)^{D}}.  
\label{Psi_function}
\end{equation}
Here $D$ is the system dimensionality, $\mathbf{G}\left( \mathbf{r}-\mathbf{r}^{\prime }|t\right)$ is 
the $2\times 2$ Green function for the total Hamiltonian in Eq.(\ref{Hamiltonian}) {(see Ref. \cite{Flatte} 
for an example)}, and $\mathbf{G}\left(\mathbf{k}|t\right)$ and ${\bm\Psi }(\mathbf{k}|0)$
are the corresponding Fourier components. 
During the measurement time $T$, the von Neumann pointer, that is the coordinate
of the particle, provides information about the motion of its spin components.
Here the mapping onto the coordinate motion makes visible otherwise hidden spin dynamics over
all possible Feynman paths in the spin subspace.
%Spin-orbit coupling entangles spin and orbital dynamics in the process of spin measurement. 
%In the absence of spin-orbit coupling $\mathbf{G}\left( \mathbf{r}-\mathbf{r}%
%^{\prime }|t-t^{\prime }\right) $ determines simple spin rotation due to the 
%Zeeman $\Delta \left( \mathbf{b}{\bm \sigma}\right)/2$ term in the Hamiltonian and its
%spatial and spin structures are independent.

The rest of the paper is organized as follows. In Section 2 this approach 
will be applied to a simultaneous  measurement of
noncommuting spin components for a two-dimensional particle with the 
Rashba spin-orbit coupling. We will show why the attempt of 
instantaneous measurement fails, and that the actual measured quantities are the
averaged over the measurement time spin components. In Section 3 we study the von Neumann measurement
and {measurement-induced dephasing}
of a spin rotating in a magnetic field for a one-dimensional particle
with non-commuting spin-orbit and Zeeman terms. In Section 4 we summarize the results and show possible
extension and generalizations of the relation between spin-orbit coupling and spin measurement procedures.  

\section{Measurement of two noncommuting spin components}

\subsection{Measurement, Feynman paths, and outcome}

The Heisenberg uncertainty relation established the limit
on precision of instantaneous measurement of observables corresponding to two
non-commuting operators in terms of the expectation value of their
commutator. Although this general statement is one of the basic properties of  
quantum motion, the  measurement procedure itself is 
still an unresolved issue \cite{AK,SHE,SP0,SP9,SP4,SP5}
even for momentum and coordinate observables.
Arthurs and Kelly \cite{AK} considered two meters
employed to measure jointly particle's position and momentum. The 
analogue of the Arthurs-Kelly experiment for non-commuting spin components
was proposed in Ref. \cite{SP0}. An implementation of such a measurement
through coupling to radiation modes was considered in Ref. \cite{SP4} and through
coupling of quantum spin to classical Ising states, in Ref. \cite{SP5}.
However, the straightforward interpretation of proposed experiments is
hardly possible. In this section we remind the reader the measurement scheme
using spin-orbit coupling \cite{Sokolovski}, where 
coordinate pointers are attached to non-commuting spin components. For this
purpose we return to the Hamiltonian (\ref{Hamiltonian}) and study spin
dynamics of a single wavepacket in the absence of external magnetic field.

Without loss of generality, we choose the spin-orbit coupling in the
conventional Rashba form. Neglecting the internal dynamics of the pointer
(exact condition will be given further in the text) we obtain
the Schr\"{o}dinger equation 
\begin{equation}
i\partial _{t}{\bm\Psi }(x,y|t)=-i\widetilde{\alpha}(\partial
_{x}\sigma _{y}-\partial _{y}\sigma _{x}){\bm\Psi }(x,y|t),
\label{Schroedinger}
\end{equation}
with an initial condition: 
\begin{equation}
{\bm\Psi }(x,y|0)=\psi_{0}(x,y){\bm\xi },  \label{psi_0}
\end{equation}
where ${\bm\Psi }(x,y|t)$ and ${\bm\xi }$ are two-component
spinors (we will usually employ the representation of $\sigma_{z}-$ eigenstates), 
$\psi_{0}(x,y)$ is the initial wavepacket, and $\widetilde{\alpha}\equiv\alpha/\sqrt{2}$. 
%The evolution of
%the initial state is given by: 
%\begin{equation}
%{\bm\Psi}(x,y|t)=\exp(-iHt){\bm\Psi}(x,y|0).
%\label{exp_Ht}
%\end{equation}
Equation (\ref{Schroedinger}) describes spin 1/2 coupled to
two von Neumann pointers \cite{vN} with positions $x$ and $y$, respectively,
in attempt to measure  spin components $\sigma_{y}$ and 
$\sigma_{x}$ simultaneously. 

Considering first measurement of a single spin component, $\sigma_{y}$, by choosing 
for a moment $H_{\rm so}=-i\widetilde{\alpha}\partial_{x}\sigma_{y}$, 
offers a useful insight. First, we present initial state as
${\bm\Psi}(x,y|0)=\psi_{0}(x,y)\left(\zeta_{1}|1\rangle_{y}+\zeta_{2}|-1\rangle_{y}\right),$
where $|1\rangle_{y}$ and $|-1\rangle_{y}$ are the eigenstates of $\sigma_{y}$ 
with the corresponding eigenvalues.  
The operator $\exp(-\widetilde{\alpha} t\partial_{x}\sigma_{y})$
splits this initial state into two components traveling
along the $x$-axis with opposite speeds $\widetilde{\alpha}$ as 
$\zeta_{1}\psi_{0}(x-\widetilde{\alpha}t,y)|1\rangle_{y}+\zeta_{2}\psi_{0}(x+\widetilde{\alpha}t,y)|-1\rangle_{y}.$
If the wave packets are well separated, and the particle is found at a location $x$,  
$x/\widetilde{\alpha}t$ approximates the value of $\sigma_y=\pm1.$ The accuracy 
of the approximation depends on the width of the $\psi_{0}(x,y)$ - for a very narrow initial distribution
we would only have $x\approx\widetilde{\alpha}t$ or $x\approx-\widetilde{\alpha}t$, realizing 
a conventional von Neumann measurement of a single spin component. 

However, our case is more complicated. Since $\sigma _{x}$ and $\sigma _{y}$, and, thus, $v_{y}$ and $%
v_{x} $, do not commute, the pointer in Eq.(\ref{Schroedinger}) does not have
a well defined two-component velocity. To study its motion, we slice the
time interval $[0,T]$ into $L$ subintervals $\epsilon =T/L$, and take the limit by 
the Lie-Trotter formula \cite{TROT} 
\begin{equation}
\exp [-\widetilde{\alpha} T(\partial _{x}\sigma _{y}-\partial _{y}\sigma
_{x})]=\lim_{L\rightarrow \infty }[\exp (-\widetilde{\alpha} \epsilon \partial
_{x}\sigma _{y})\exp (\widetilde{\alpha} \epsilon \partial _{y}\sigma _{x})]^{L},
\label{Trotter}
\end{equation}
where 
\begin{equation}
\exp (\pm \widetilde{\alpha} \epsilon \partial _{j}\sigma _{i})=\sum_{m=\pm 1}|m{%
\rangle }_{i}\exp (\pm m\widetilde{\alpha} \epsilon \partial _{j})_{i}{\langle }m|
\label{decomposition}
\end{equation}
with $\sigma _{i}|m{\rangle }_{i}=m|m{\rangle }_{i}$ and $i=x,y$,
corresponding to all possible, hidden in the absence of spin-orbit coupling,
virtual Feynman walks on an infinite lattice $x(j_{x})=j_{x}\widetilde{\alpha}
\epsilon $, $y(j_{y})=j_{y}\widetilde{\alpha} \epsilon $, $j_{x},j_{y}=\ldots
-1,0,1,\ldots $ reminiscent of the Feynman's checkerboard for a Dirac electron 
\cite{Feyn}. 
\begin{figure}[tb]
\begin{center}
\includegraphics*[width=0.85\textwidth]{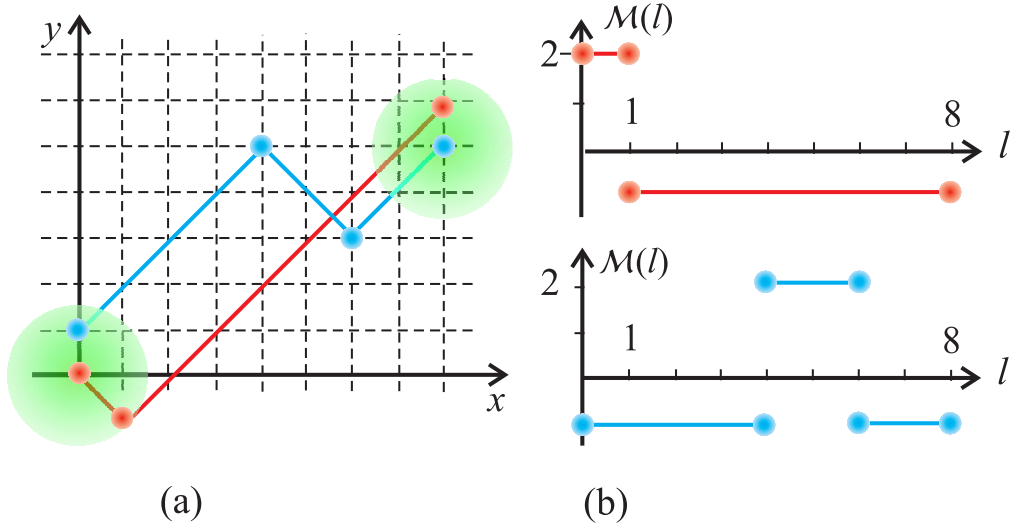}
\end{center}
\caption{(a) Feynman paths for two possible particle's
virtual histories in the $xy$-plane ($L=8$). Both initial and final positions
are unresolved being located under the Gaussian (green circles) of the width $w$. 
{We consider a short measurement where $w$ remains constant for $t<T$.}  (b) The
time dependence of $\mathcal{M}(l)$ function in Eqs.(\ref{Ml1}) and (\ref{Ml2}) 
for these two paths. For both paths $\left\langle \sigma _{y}\right\rangle
_{T}=1$. For the upper path $\left\langle \sigma _{x}\right\rangle _{T}=-3/4$%
, and for the lower path $\left\langle \sigma _{x}\right\rangle _{T}=-1/2.$}
\label{fig:FIG1}
\end{figure}
In every time step the particle moves forwards or backwards along the $x$-
and $y$-axis, and its position at the end of the measurement, $t=T$, is
determined by the differences, $\Delta n_{x}$ and $\Delta n_{y}$, between
the numbers of forward and backward steps taken in each direction or, more
precisely, by the interference between all paths sharing the same $%
\Delta n_{x}$, and $\Delta n_{y}$ (see Fig.\ref{fig:FIG1}(a)). Next we
assign values $m_{i}({l})=\pm 1$, ${l}=1,\ldots ,L$ to $\sigma _{i}$ in each step and define: 
\begin{equation}
\Delta n_{i}=\sum_{l=1}^{L}m_{i}({l}).
\label{mean_values}
\end{equation}
As a result, finding the pointer at a location $(x,y)$ determines 
\textit{time averages} of the spin components, ${\langle }\sigma _{x}{%
\rangle }_{T}$ and ${\langle }\sigma _{y}{\rangle }_{T}$, defined for the
spin-space Feynman paths (Fig.\ref{fig:FIG1}(b)) as: 
\begin{equation}
{\langle }\sigma _{i}{\rangle }_{T}\equiv \frac{1}{T}\int_{0}^{T}\sigma_{i}(t)dt,  \label{mean_values_1}
\end{equation}
to an accuracy determined by the position spread of the initial $\psi_{0}(x,y)$. 
In this Figure, we characterize both spins by a single function 
\begin{eqnarray}
&&\hspace{-2cm}\mathcal{M}(l)=2,\mbox{ if }m_{x}(l)=m_{y}(l)=1,\quad 
\mathcal{M}(l)=1,\mbox{ if }m_{x}(l)=1,m_{y}(l)=-1 
\label{Ml1}\\
&&\hspace{-2cm}\mathcal{M}(l)=-1,\mbox{ if }m_{x}(l)=-1,m_{y}(l)=1,\quad 
\mathcal{M}(l)=-2,\mbox{ if }m_{x}(l)=m_{y}(l)=-1.
\label{Ml2}
\end{eqnarray}
The inverse is given by: $m_{x}(l)=\mbox{sign}{\mathcal{M}(l)}$, and $m_{y}(l)=(-1)^{\mathcal{M}{(l)}}\mbox{sign}{\mathcal{M}(l)}$.

\subsection{Transition amplitudes and expectation values of observables}

To {study the details of the measurement procedure, we begin with introducing} 
the radius $R_{\mathrm{so}}\equiv\widetilde{\alpha}T$ 
and note that a particle initially localized at the origin
would never leave the 'allowed' circle $r\equiv(x^{2}+y^{2})^{1/2}\le R_{\mathrm{so}}$. 
To study the coupled evolution, we define the initial state as the spatial Gaussian,
corresponding to the particle release from the ground state of a harmonic potential:
\begin{equation}
{\bm\Psi}(x,y|0)=\frac{\sqrt{2}}{\sqrt{\pi }w}\exp (-r^{2}/w^{2})\left[ 
\begin{array}{l}
\xi _{1} \\ 
\xi _{2}
\end{array}
\right],  \label{Gaussian}
\end{equation}
and the matrix $\mathbf{U}(x,y|T;\psi_{0})$, such that at the end of the measurement 
(in the following we omit the explicit dependence on the initial state $\psi_0(x,y)$)
\begin{equation}
{\bm\Psi} (x,y|T)={\mathbf{U}}(x,y|T)\left[ 
\begin{array}{l}
\xi _{1} \\ 
\xi _{2}
\end{array}
\right].
\label{PsiT}
\end{equation}
Using Eqs.(\ref{expHt}) and (\ref{Psi_function}) for
the Hamiltonian corresponding to Eq.(\ref{Schroedinger}),   
we find that in the cylindrical coordinates $\mathbf{U}(r,\varphi|T)$ is a
Hermitian matrix with 
\begin{eqnarray}
&&\hspace{-2cm}U_{11}(r|T)=U_{22}(r|T)=w\int_{0}^{\infty }\exp \left(
-k^{2}w^{2}/4\right) \cos \left( R_{{\mathrm{so}}}k\right) J_{0}(kr)\frac{kdk%
}{\sqrt{2\pi }}, 
 \label{U_ij1}\\
&&\hspace{-2cm}U_{12}(r,\varphi|T)={U_{21}}^{*}(r,\varphi|T)=ie^{-i\varphi
}w\int_{0}^{\infty }\exp \left( -k^{2}w^{2}/4\right) \sin \left( R_{{\mathrm{%
so}}}k\right) J_{1}\left( kr\right) \frac{kdk}{\sqrt{2\pi }},  \label{U_ij2}
\end{eqnarray}
where $J_{n}(z)$ is the Bessel function of the first kind of order $n$ and $%
\varphi $ is the angle between $\mathbf{r}$ and the $x$-axis \cite{Dugaev}. 
As a results, the initial state shows a two-dimensional spread and after the measurement 
produces the density ${\bm\Psi}^{\dagger}(r,\varphi|T){\bm\Psi}(r,\varphi|T)=|\Psi_{1}^{2}(r,\varphi|T)|+|\Psi_{2}^{2}(r,\varphi|T)|,$
concentrated in a ring of a radius $R_{\mathrm{so}},$ width $\approx w$ 
and dependent on the angle $\varphi.$  The large argument
asymptotes of the Bessel functions \cite{ABRAM} 
\begin{equation}
J_{n}(kr)\sim \sqrt{\frac{2}{\pi kr}}\cos (kr-n\pi /2-\pi /4)  
\label{Bess_amp}
\end{equation}
show that in the limit $w\rightarrow 0$ the integrals become singular as $r$
approaches $R_{{\mathrm{so}}}$, and the measurement is possible only with a
finite accuracy at nonzero width. 

It is worth mentioning that an additional obstacle to a highly accurate
measurement appears due to a finite mass of the particle, where 
packet spreads with the characteristic speed of the order of $v_{\rm sp}=1/wM.$ 
Thus, the conditions of the precise measurement can be formulated as: (i) $\alpha \gg v_{\rm sp},$ to assure that the 
wavepacket dynamics is due to the spin-orbit coupling, not due to the 
broadening resulting from the finite $M$,  and (ii) $\alpha T\gg w$ to assure that 
the split of the wavepacket at time $T$ is sufficient to perform the measurement.  

Three implementations of this procedure can be considered.

1) For electron in semiconductor structures with typical $\alpha \sim 10^{6}$ cm/s,
{that is, in conventional units, $\hbar\alpha\sim 10 \mbox{ meVnm},$}
and $M\sim 5\times 10^{-29}$ g, condition $\alpha \gg v_{\rm sp}$ can be
achieved for $w>100$ nm, and corresponding $T>10^{-11}$ s.

2) For topological insulators, $\alpha \sim 10^{8}$ cm/s ($\hbar\alpha\sim 1 \mbox{ eVnm},$) 
with infinite $M$, and, therefore, the first condition $\alpha \gg
v_{\rm sp}$ is always satisfied. However, at the moment, it is
difficult to prepare and control individual electron states  
is these systems.   

3) For cold bosonic and fermionic atoms, taking 
$\alpha \sim 10$ cm/s ($\hbar\alpha\sim 10^{-6} \mbox{ K}\mu\mbox{m},$ with the relevant energy 
scale measured in Kelvin and the distance in micron) 
and fermion $^{40}$K as an example, we obtain $w>10^{-4}$ cm and $T>10^{-5}$ s. 
{We mention here that coherent 
many-particle spin-orbit coupled Bose-Einstein condensates with the pseudospin 
1/2 such as $^{87}\mbox{Rb}$ can be an interesting system for the proposed measurement.  
Condensates with a weak effective interatomic interaction are preferable for this purpose since 
with the increase in this interaction, the wave packet spread may be controlled 
by the interatomic repulsion or attraction rather than by the initial width. As a result, 
the corresponding spreading rate should be taken in the above criteria of accurate measurement.}

\begin{figure}[tb]
\begin{center}
\includegraphics*[width=0.6\textwidth]{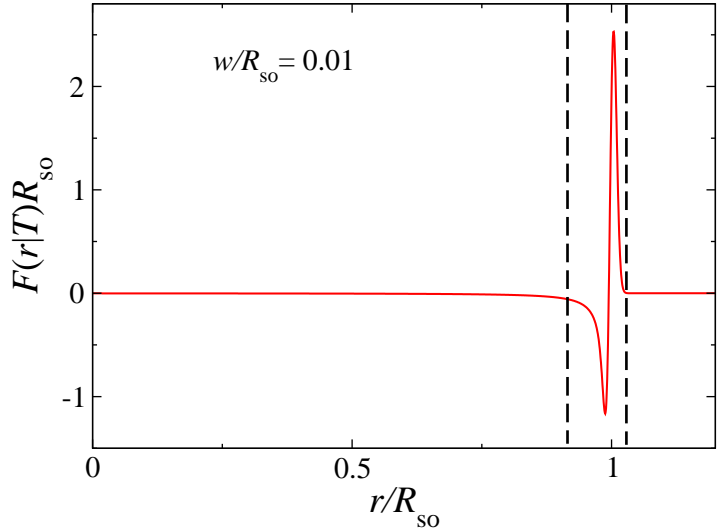}
\end{center}
\caption{The radial function $F(r|T)$ for $w/R_{\mathrm{so}}=0.01$. Vertical dashed lines
mark the interval (ring) containing most of the probability density.}
\label{fig:FIG2}
\end{figure}

For a highly accurate measurement with $w\ll R_{\mathrm{so}}$ main contributions to the integrals in 
(\ref{U_ij1}),(\ref{U_ij2}) come from the domain where $1/R_{\mathrm{so}}\ll k \sim 1/w.$ Replacing
the Bessel functions by their asymptotes (\ref{Bess_amp}), and neglecting
contribution of rapidly oscillating ($\sim\cos(R_{\mathrm{so}}k)$) terms in the integrand yields 
\begin{eqnarray}
&&\mathbf{U}(r,\varphi|T)\sim F(r|T)\left[\sigma_{0}+\sigma_{x}\sin\varphi - \sigma_{y}\cos\varphi\right], 
 \label{sigtheta}\\
&&F(r|T)=\frac{w}{\sqrt{R_{\mathrm{so}}}}\int_{0}^{\infty }\exp \left(
-k^{2}w^{2}/4\right) \cos [(r-R_{{\mathrm{so}}})k+\pi /4]k^{1/2}
\frac{dk}{2\pi},  \label{U_ij_1}
\end{eqnarray}
demonstrating that spin-coordinate entanglement is, in fact, spin-angle entanglement. 
For a small $w$, the radial function $F(r|T)$, obtained in Ref.\cite{Sokolovski}, has
maximum and a minimum close to $r=R_{\mathrm{so}}$, rapidly decreases for $%
r>R_{\mathrm{so}}$, and exhibits a somewhat slower decay for $r<R_{\mathrm{so}}$, as shown in Fig.\ref{fig:FIG2}. 
Figure \ref{fig:FIG1} corresponds to a less accurate
measurement, where $R_{\mathrm{so}}$ and $w$ are of the same order of magnitude. 

Although the probability density is concentrated in a narrow vicinity of $R_{\rm so}$, 
the broad angular distribution in Eq.(\ref{sigtheta}) appeared 
as a result of the measurement, has a relatively broad distribution of moments: 
\begin{eqnarray}
&&\langle x(T)\rangle = -R_{\rm so}\langle\sigma_{y}(0)\rangle/2, \qquad \langle y(T)\rangle = R_{\rm so}\langle\sigma_{x}(0)\rangle/2; \\
&&\langle x^2(T)\rangle = R_{\rm so}^{2}/2, \qquad \langle y^2(T)\rangle = R_{\rm so}^{2}/2,
\end{eqnarray}
where $\langle\sigma_{y}(0)\rangle$ and $\langle\sigma_{x}(0)\rangle$ are expectation values of corresponding spin components at $t=0$. 
The $\varphi-$dependent spin orientations at $t=T$ were presented in Ref.\cite{Sokolovski}. 

To conclude this section, switching a Rashba coupling over a short time $T$
simulates a von Neumann measurement of two non-commuting spin components. Here, 
the particle itself plays the role of a pointer which correlates
its position, $(x,y)$, with the time average of the corresponding spin
components, ${\langle }\sigma _{x}{\rangle }_{T}$ and ${\langle }\sigma _{y}{%
\rangle }_{T}$ evaluated along Feynman paths defined for the two spin
variables (Fig.\ref{fig:FIG1}). There are infinitely many trajectories which share the
same ${\langle }\sigma _{x}{\rangle }_{T}$ and ${\langle }\sigma
_{y}{\rangle }_{T}$ and lead to the same pointer position. We stress here that
since these time averages are not instantaneous values of the 
spin components, they can be measured simultaneously to any
desired accuracy. Indeed, the Feynman paths shown in Fig.\ref{fig:FIG1} have
no intrinsic time scale and can be cut into arbitrarily small $\epsilon
/T\rightarrow 0$ pieces. Thus, even in the pulse limit $T\rightarrow 0$,
under condition $\widetilde{\alpha} T=\mathrm{const}$ one does not attain unique
instantaneous values of the two spin components: no matter how short is $T$,
all paths shown in Fig.\ref{fig:FIG1}(a) contribute to the transition
amplitude (\ref{PsiT}), densely filling the area with $r\le R_{\rm so}$, 
and no measurement can catch $\sigma _{x}$ and $%
\sigma _{y}$ simultaneously. Thus, the fact that the particle can choose an infinite
set of completely different Feynman paths at any short measurement time 
forbids us to define sharp instantaneous values for the non-commuting spin components.

%It is worth noting that recent developments in producing synthetic
%spin-orbit coupling by optical fields allow one to realize the three-dimensional spin-orbit
%coupling \cite{Anderson} in the form $\left(\cdot{\bm\sigma}{\bf k}\right)$ and, therefore, for an
%attempt of von Neumann measurement of all three spin components in a single experiment providing
%another realization of full qubit monitoring \cite{Ruskov}.

\section{Measurement of spin in magnetic field}

\subsection{Measured quantity, Feynman paths, and Green functions}

Next we consider measurement {of a spin of a particle in a one-dimensional} system, where the
velocity operator and the Zeeman term do not commute. Such a measurement can be realized in semiconductor
quantum wires \cite{wire} and narrow waveguides for the Bose-Einstein condensates \cite{plaja,meyrath}
with implemented spin-orbit coupling. We take the one-dimensional version of 
Hamiltonian (\ref{Hamiltonian}) as:
\begin{equation}
\hat{H}_{\rm 1D}=\frac{k^{2}}{2M}+\alpha k\sigma_{z}+\frac{\Delta }{2}\left(\mathbf{b}\cdot{\bm\sigma }\right),  
\label{Hamiltonian_1D}
\end{equation}
and assume that at $t=0,$ ${\bm\Psi}(x|t)=\psi_{0}(x){\bm \xi}$, and $\mathbf{b}=(\sin\theta,0,\cos\theta).$
We neglect the diamagnetic effects of vector potential ${\cal A}_{x,y}$ in a one-dimensional system.
In addition, we mention that for cold atoms \cite{Wang10,Stanescu,Lin,Wang,Sommer,Cheuk,Liu} 
a synthetic Zeeman-like coupling can be realized without diamagnetic terms.  
{Here, the spin-orbit coupling term $\alpha k\sigma_{z}$ satisfies the condition 
of the von Neumann measurement and the resulting dynamics simulates the quantum measurement \cite{Sokolovski_mes} 
of a single spin component, $\sigma_{z}$ in this case.} Despite apparent simplicity, this system demonstrates
rather nontrivial behavior, presented below. 

Although, unlike in the previous Section, only one component, is measured, 
the spin is not static, but undergoes a precession due to the Zeeman term in (\ref{Hamiltonian_1D}).
With the SOC turned on, each component of the particle's momentum adds an additional field along the $z$-axis.
As a result, the  spin moves in a magnetic field, whose direction and magnitude are {\it fundamentally} uncertain.
The velocity term $\alpha\sigma_{z}$ {given by the commutator $i[\hat{H}_{\rm 1D},x]$,} 
does not commute with the Hamiltonian, and wavepacket nonuniformly spreads even 
if  the  particle's  mass is infinite. 
We begin with the general properties of this dynamics and then discuss expectation values of observables 
in relation to the von Neumann measurement procedure. 

\begin{figure}[tb]
\begin{center}
\includegraphics*[width=0.85\textwidth]{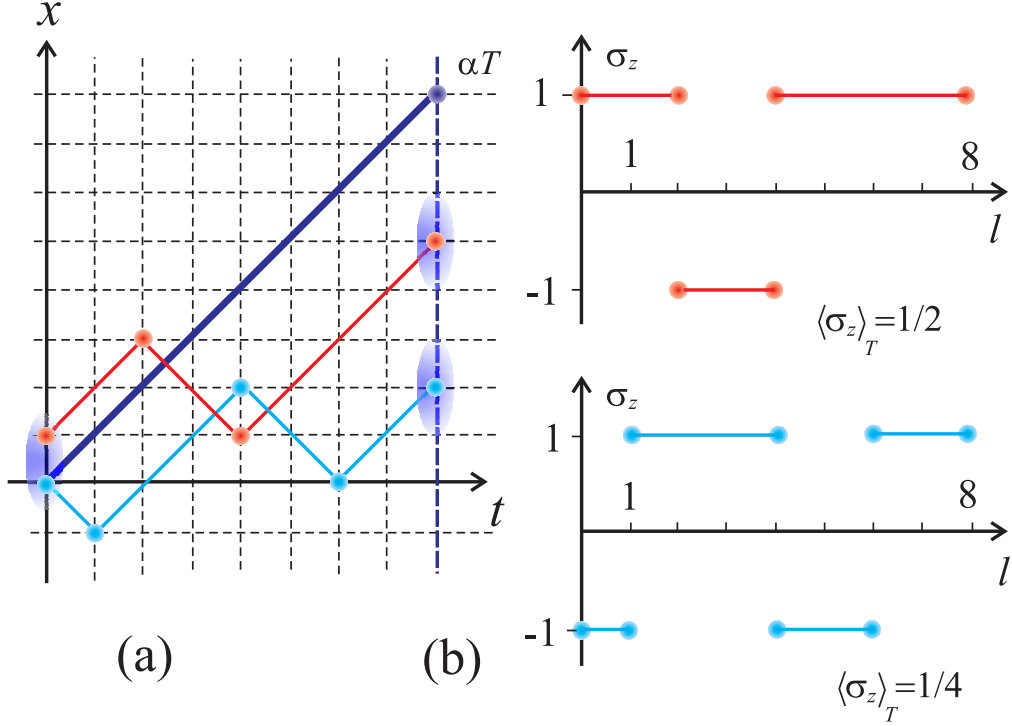}
\end{center}
\caption{(a) Feynman checkerboard in the $(x|t)$ space for two possible
virtual paths (we take $L=8$). Initially unresolved positions under the
one-dimensional Gaussian of the widths $w$ (shown as the elongated ellipses)
become resolved at the end of the measurement at $t=T$. The bold straight
line corresponds to one of the possible paths $x=\pm\alpha t$ in the absence of a magnetic
field. (b) The time dependence of $\sigma _{z}$ corresponding to the
particle displacement for these two paths.}
\label{fig:FIG3}
\end{figure}

Following the approach from the previous Section,
and neglecting for a moment the kinetic energy, we slice the interval $[0,T]$ into $L$ sub-intervals 
of a length $\epsilon$, apply the Lie-Trotter formula now to 
$\exp\{-i\epsilon[\alpha \hat{k}\sigma _{z}+\Delta\left( \mathbf{b}\cdot{\bm\sigma}\right)/2]\}$, 
and introduce virtual Feynman paths where $\sigma_z$
takes a value $s_j=\pm1$ at each $\epsilon$-interval, $j=1,2,\ldots L$. 
The exact propagator can be expressed as a path sum. Taking $L\rightarrow\infty$ we obtain 
\begin{equation}\label{22}
\exp\left(-i\hat{H}_{\rm 1D}T\right)=\sum_{\rm paths}\hat{U}_{\rm 1D}[{\rm path}],
\end{equation}
where the operator
\begin{eqnarray}\label{23}
\hspace{-2cm}\hat{U}_{\rm 1D}[{\rm path}]=\exp(-\alpha T{\langle }\sigma _{z}{\rangle }_{T}\partial_x)\quad \quad\\ \nonumber
%\hat{u}[path]=
\hspace{-2cm}\times \lim_{L\to \infty}
|s_{L}\rangle\langle s_{L}|\exp[-i\epsilon\Delta\left(\mathbf{b}\cdot{\bm\sigma}\right)/2]|s_{L-1}\rangle
\ldots
\langle s_2|\exp[-i\epsilon\Delta\left(\mathbf{b}\cdot{\bm\sigma}\right)/2]|s_{1}\rangle\langle s_1|,
\end{eqnarray}
translates the initial state by a distance $\alpha{\langle}\sigma_{z}{\rangle}_{T}T$. 
If $\psi_{0}(x)$ is completely localized, $|\psi_{0}^{2}(x)|=\delta(x)$, 
from the final position of the pointer $x$ one can accurately deduce the value 
of $\langle\sigma_{z}\rangle_{T}=x/\alpha T$. 
A spread in the initial positions leads to a finite error in the determination 
of $\langle\sigma _{z}\rangle_{T}$. 
Finally, including the pointer kinetic energy leads to spreading of the initial 
wave packet $\psi_0(x)$ in the measurement time.  
The corresponding Feynman checkerboard in the $(t,x)$ space mapping spin 
motion on spatial dynamics is shown in Fig.\ref{fig:FIG3}(a).
%As in Section 2, we adopt, when necessary, the Gaussian choice for the 
%initial wave function in the coordinate and momentum spaces: 
%\begin{equation}
%\psi_{0}(x)=\frac{1}{\sqrt[4]{\pi w^{2}}}e^{-x^{2}/2w^{2}},
%\qquad \psi_{0}(k)=\left( 2\sqrt{\pi }w\right) ^{1/2}e^{-k^{2}w^{2}/2},
%\label{psi_p}
%\end{equation}
%using the same notation in  both representations, and the initial spin state: 
%\begin{equation}
%\xi_{1} =\cos\left(\beta/2\right)e^{i\phi}, \quad  
%\xi_{2} = \sin\left(\beta/2\right).
%\label{initial_state}
%\end{equation}  

Although the Lie-Trotter formula is valuable for the understanding of the
virtual Feynman paths, Eq.(\ref{Psi_function}) is more convenient for detailed 
calculations, similar to the approach used in Section 2. 
Introducing notations $\gamma \equiv\Delta\cos\theta/\alpha,$
$q\equiv 2k+\gamma$, $C\equiv \cos \left( t\sqrt{q^{2}\alpha ^{2}+%
\widetilde{\Delta }^{2}}/2\right),$ $S\equiv \sin \left( t\sqrt{q^{2}\alpha
^{2}+\widetilde{\Delta }^{2}}/2\right),$ where $\widetilde{\Delta}\equiv{\Delta}\sin
\theta,$ and {using Eq.(\ref{expHt})} with 
Hamiltonian (\ref{Hamiltonian_1D}), we obtain the 
Green function in Eq.(\ref{Psi_function}) in the form:
\begin{eqnarray}
\hspace{-2cm}{G}_{11}(x|t) &=&e^{-i\gamma x/2}\int_{-\infty}^{\infty} \left[ C-i\frac{q\alpha }{%
\sqrt{q^{2}\alpha ^{2}+\widetilde{\Delta }^{2}}}S\right]
e^{iqx/2}e^{-i\left( q/2-\gamma /2\right)^{2}t/2M}\frac{dq}{%
4\pi },  \label{g11}\\
\hspace{-2cm}{G}_{22}(x|t) &=&e^{-i\gamma x/2}\int_{-\infty}^{\infty} \left[ C+i\frac{q\alpha }{%
\sqrt{q^{2}\alpha ^{2}+\widetilde{\Delta }^{2}}}S\right]
e^{iqx/2}e^{-i\left( q/2-\gamma /2\right)^{2}t/2M}\frac{dq}{%
4\pi },  \label{g22}\\
\hspace{-2cm}{G}_{12}(x|t) &=&{G}_{21}(x|t)=-ie^{-i\gamma x/2}\widetilde{\Delta}
\int_{-\infty}^{\infty} \frac{S}{\sqrt{q^{2}\alpha ^{2}+\widetilde{\Delta }^{2}}}%
e^{iqx/2}e^{-i\left( q/2-\gamma /2\right)^{2}t/2M}
\frac{dq}{4\pi}.
\label{g12}
\end{eqnarray}

We begin with two simple limiting cases.

% taking $v_{\mathrm{\rm sp}}=0$ for simplicity.

Case 1. The spin-orbit coupling and Zeeman terms commute, with $\theta =0$ as an example. The Green function 
\begin{equation}
\hspace{-2cm}\mathbf{G}\left(x-x^{\prime}|T\right)=
\left[ 
\begin{array}{ll}
e^{-iT\Delta/2} g_{0}(x-x^{\prime}-\alpha T|T) & 0 \\ 
0 & e^{iT\Delta/2} g_{0}(x-x^{\prime}+\alpha T|T)
\end{array}
\right],  \label{G_diag}
\end{equation}
is diagonal in the spin space. Here $g_{0}(x|t)=(M/2\pi i t)^{1/2}\exp(iMx^2/2t)$
is the free particle propagator \cite{Feyn,Shankar}, which in the limit 
of infinite $M$ tends to the Dirac $\delta(x)-$function. For a  nonzero $\sin\theta$, 
Green function (\ref{G_diag}) is valid for short time $t\ll1/\widetilde{\Delta},$
where the effect of magnetic field on the spin precession is still weak.   
%As discussed in the Section 3, if the spin-dependent
%terms in the Hamiltonian commute, the dynamics is the splitting
%of the initial wave packet, meaning that for infinitely heavy particle and $\Delta=0$: 
%\begin{equation}
%{\bm\Psi }(x|t)=\xi_{1}\psi_{0}(x-\alpha t)|1\rangle_{z}+\xi_{2}\psi_{0}(x+\alpha t)|-1\rangle_{z},  
%\label{psi_x_t}
%\end{equation}
%producing entangled spin-coordinate state and defining the accuracy of spin measurement, as
%corresponds to the paths $x=\alpha t$ shown in Fig.\ref{fig:FIG3}(a) and $x=-\alpha t,$ not
%shown there. The off-diagonal components of the spin density matrix $\hat{{\bm \rho}}$ 
%vanish as the overlap of $\psi_{0}(x-\alpha t)$ and $\psi_{0}(x+\alpha t)$
%decreases, corresponding to the formation of a mixed state resulting from the measurement. 

Case 2. Without  spin-orbit coupling, $\alpha =0,$ the Green function
factorizes into a free propagator in the coordinate space, and a spin part describing Larmor precession:
\begin{equation}
\mathbf{G}\left(x-x^{\prime}|T\right)=
\exp{\left[-i\frac{\Delta}{2}(\mathbf{b}\cdot{\bm\sigma})T\right]} g_{0}(x-x^{\prime}|T).
\end{equation}
No spin measurement can be done here: although all virtual Feynman paths 
in Fig.\ref{fig:FIG3}(b) are possible and interfere, they cannot be mapped
on the $x-$coordinate motion. 

\begin{figure}[tb]
\begin{center}
\includegraphics*[width=0.65\textwidth]{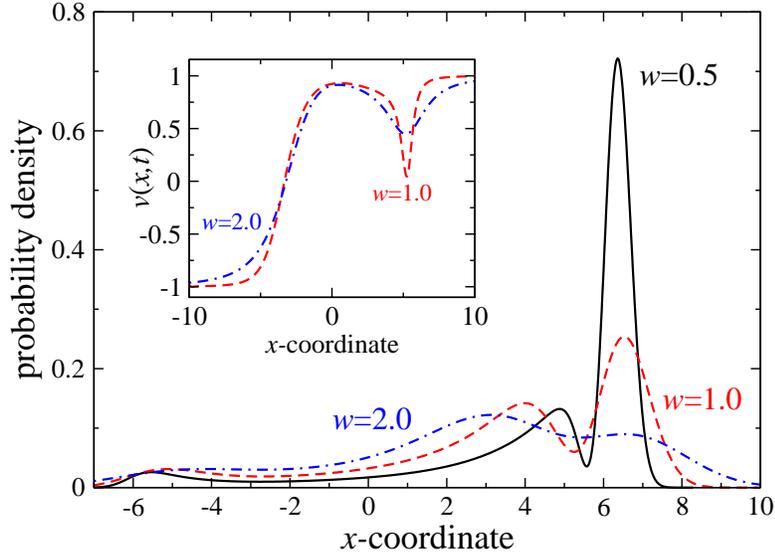}
\end{center}
\caption{Total density as a function of coordinate at $t=2\pi$ for packet widths $w$ 
shown in the plot. The inset shows the coordinate
distribution of velocity for $w=1.0$ and $w=2.0$. 
Similar dip at coordinate $r$ close
to $R_{\rm so}$ resulting from interference of different paths,
is seen in calculations for two-dimensional walks, as shown in Fig.\ref{fig:FIG2} \cite{Sokolovski}.
We use here units $1/\widetilde{\Delta}$ for time, $\alpha/\widetilde{\Delta}$ for length,
$\theta=\pi/4$, and $\xi_{1}=1$.
}
\label{fig:FIG4}
\end{figure}

In the general case of non-commuting spin-orbit and Zeeman terms, the particle can choose all the paths
shown in Fig. \ref{fig:FIG3}(a), where 
$x(T)-x(0)=\alpha\left\langle \sigma_{z}\right\rangle_{T},$ same as discussed for the two-dimensional motion.
This dynamics is complicated due to spin precession in magnetic
field, leading to nonzero sum of contributions from different paths.
The precession stops once a steady spin state with $\langle\sigma_{y}(\infty)\rangle =0$ is reached, although the 
wavepacket continues to move and to spread. 

\subsection{Finite-time measurement and coupled spin-coordinate evolution}

To focus on the effect of spin-orbit coupling, we consider 
(if not explicitely  stated otherwise)  the particle 
of  infinite mass, avoiding the packet broadening due its initial kinetic energy.
The infinite mass condition is essentially the first requirement of a precise measurement,
$\alpha \gg v_{\rm sp}$, formulated in Section 2. To illustrate the origin of the complicated 
character of the spin dynamics even in the infinite mass limit, we present the continuity equation:  
\begin{equation}
\frac{\partial \rho(x|t) }{\partial t}+\frac{\partial j(x|t)}{\partial x}=0,
\end{equation}
where $\rho(x|t)={\bm\Psi}^{\dagger}(x|t){\bm\Psi }(x|t)$ is the probability 
density, and $j(x|t)={\bm\Psi}^{\dagger}(x|t)\sigma _{z}{\bm\Psi }(x|t)$ is the spin-determined
flux. The resulting local velocity 
\begin{equation}
v(x|t)=\frac{{\bm\Psi}^{\dagger}(x|t)\sigma_{z}{\bm\Psi }(x|t)}
{{\bm\Psi}^{\dagger}(x|t){\bm\Psi}(x|t)}  
\label{continuity}
\end{equation}
strongly varies with time and coordinate leading, it turn, to the nontrivial dependence
of probability and current densities.

\begin{figure}[tb]
\begin{center}
\includegraphics*[width=0.65\textwidth]{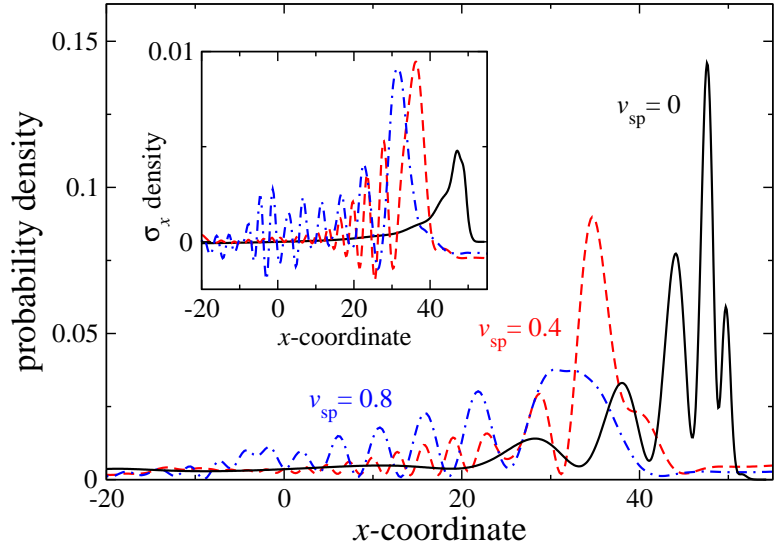}
\end{center}
\caption{Total density as a function of coordinate at $t=16\pi$ and $w=1$
for different $M$-dependent velocities of the initial state spread 
$v_{\mathrm{\rm sp}}=0,0.4,$ and 0.8 as shown near the plots. The inset shows (with
the corresponding style of the lines) the distribution of density
of $x$-component of spin ${\bm\Psi}^{\dagger}(x|t)\sigma_{x}{\bm\Psi}(x|t).$
We use units $1/\widetilde{\Delta}$ for time, $\alpha/\widetilde{\Delta}$ for length,
$\theta=\pi/4$, and $\xi_{1}=1$.}
\label{fig:FIG5}
\end{figure}

We begin with the snapshots of the probability and spin densities 
showing the role of the width of the packet as
the measurement tool. Although our approach is general, in calculations we use
(as in Section 2) the Gaussian initial wave function in the coordinate and momentum spaces: 
\begin{equation}
\psi_{0}(x)=\frac{1}{\sqrt[4]{\pi w^{2}}}e^{-x^{2}/2w^{2}},
\qquad \psi_{0}(k)=\left( 2\sqrt{\pi }w\right)^{1/2}e^{-k^{2}w^{2}/2},
\label{psip}
\end{equation}
and the initial spin state: 
\begin{equation}
\xi_{1} =\cos\left(\beta/2\right)e^{i\phi}, \quad  
\xi_{2} = \sin\left(\beta/2\right).
\label{initial_state}
\end{equation}  
We use Eq.(\ref{Psi_function}) with Hamiltonian (\ref{Hamiltonian_1D}) and corresponding Green
functions (\ref{g11})-(\ref{g22}) to obtain time and coordinate-dependent wave functions. 

Figure \ref{fig:FIG4} shows the distribution of density
and velocity for dimensionless time $t=2\pi $ and different packet width. Figure \ref{fig:FIG5} shows the
effect of packet spread, that is internal evolution of the pointer due to
the kinetic energy in the Hamiltonian. The spread of the pointer state
decreases the accuracy of the measurement since it considerable decreases the available range of momenta. If the mass
of the particle is infinite, the range of momenta is of the order of $1/w$. If the spread velocity
$v_{\rm sp}=1/wM$ is nonzero, at large $t$ the momentum spread decreases as $1/\sqrt{w v_{\rm sp}t}$, and, therefore,
fine details of the Green function become gradually smeared, and the spin measurement accuracy decreases with time
since the displacement of the particle is determined not only by the spin dynamics,
but also by the wave packet's  spreading. 
This statement can be understood with the following optical analogy. If $v_{\rm sp}=0$, the Green function 
is seen through a magnifying glass with a given resolution, smearing its finer details. For nonzero $v_{\rm sp}$, 
the 
Green function is seen through a diffraction grating with a relatively large period, 
which increases with time, smearing the details to even greater extent.

\begin{figure}[tb]
\begin{center}
\includegraphics*[width=0.85\textwidth]{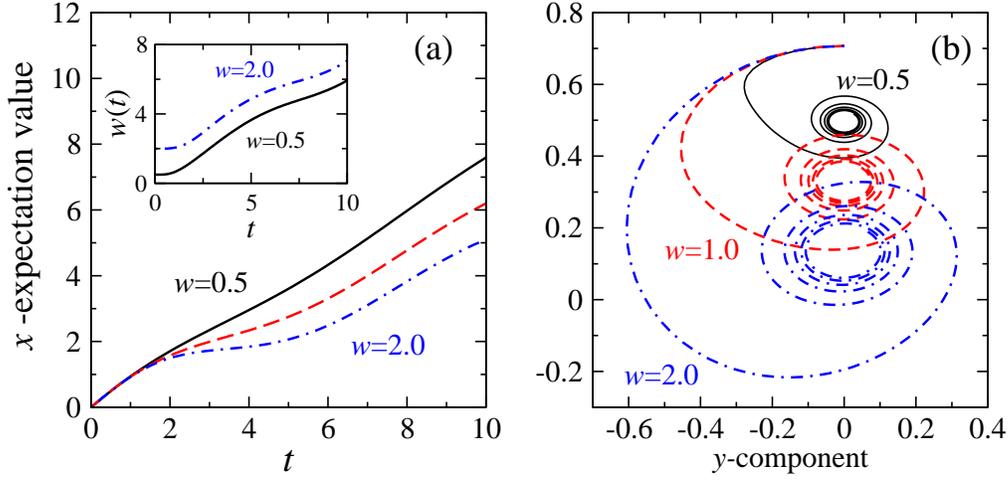}
\end{center}
\caption{(a) Time dependence of the expectation value of the
coordinate $\langle x(t)\rangle$ for different packet widths, shown near the plots. The inset shows fast broadening of
the packet. (b) The $(\langle\sigma_{y}\rangle,\langle\sigma_{\perp})\rangle$ trajectory.
We use units $1/\widetilde{\Delta}$ for time, $\alpha/\widetilde{\Delta}$ for length, 
$\theta=\pi/4$, and $\xi_{1}=1$.}
\label{fig:FIG6}
\end{figure}

To make a comparison with the analysis of the previous Section, 
we calculate the expectation values of coordinate- and spin-related observables
to see the relationship between them. We begin with tracing the following quantities: 
\begin{eqnarray}
\left\langle x(t)\right\rangle &=&\int_{-\infty}^{\infty} {\bm\Psi}^{\dagger}(x|t)x{\bm\Psi}(x|t)dx, 
\label{xt}\\
w(t) &=&\sqrt{2}\sqrt{\langle x^{2}(t)\rangle -\langle x(t)\rangle ^{2}}, 
\label{wt}
\end{eqnarray}
for the expectation value of coordinate $\langle x(t)\rangle$ and 
width of the packet $w(t)$, respectively. 

%To focus on the effect of spin-orbit coupling, we consider the particle
%of  infinite mass, avoiding the packet broadening due its initial kinetic energy.  
Figure \ref{fig:FIG6}(a) presents expectation value $\langle x(t)\rangle$ 
for different initial width of the packet and the time-dependent packet width 
in the inset. For any width
at short times we obtain $\langle x(t)\rangle=\alpha t$, when all possible paths are in the 
vicinity of the straight line in Fig.\ref{fig:FIG3}(a). After some time, dependent on $w$,
the spin explores other Feynman walks, and the dependence becomes asymptotically 
linear with the main term $\langle x(t)\rangle=\alpha \langle\sigma_{z}(\infty)\rangle t$. 
The inset shows the increase in the packet width with time, where different displacements  
$x-x^{\prime}=\alpha t\langle\sigma_{z}\rangle_{t}\le\alpha t$ are possible due to the spin precession in magnetic
field. The $x(t)$-dependences here split and the packets start to broaden at dimensionless $t$
close to $\pi$, where the spin makes a half-turn. We emphasize that the packet
broadening and nonlinear $\langle x(t)\rangle-$dependence are solely due to the 
noncommutativity of the Zeeman and the Rashba terms.

\subsection{Spin decoherence and long measurement}

To better understand the coupled coordinate- and spin dynamics we consider 
the evolution of spin components by tracing the quantities 
\begin{eqnarray}
\left\langle \sigma _{i}(t)\right\rangle &=&\int_{-\infty}^{\infty} {\bm\Psi}^{\dagger}(x|t)\sigma_{i}{\bm\Psi}(x|t)dx=
{\rm tr}\left[\hat{\bm{\rho}}(t)\sigma _{i}\right],
\label{sigmat}
\end{eqnarray}
calculated with the spin density matrix $\hat{\bm{\rho}}(t)$:
\begin{equation} 
\hat{\bm{\rho}}(t)=\int_{-\infty}^{\infty} {\bm\Psi}(x|t){\bm\Psi}^{\dagger}(x|t)dx.
\label{denmat}
\end{equation}
With the increase in $t$, the different evolution of spinor components $\Psi_{1}(x|t)$ and $\Psi_{2}(x|t)$
produces a spin-coordinate entanglement and makes the initially 
pure spin state with ${\rm tr}\hat{{\bm\rho}}^{2}(0)=1$, a mixed
one with ${\rm tr}\hat{{\bm\rho}}^{2}(t)<1$. As a result, the spin subsystem experiences a decoherence in the measurement process, and
spin moves from the initial position at the Bloch sphere $\sum_{i}\langle\sigma _{i}(0)\rangle^{2}=2{\rm tr}\hat{{\bm\rho}}^{2}(0)-1,$ 
to the inner part of the corresponding Bloch ball, where this sum is less than one.
The {spin-dependent velocity in a system with Hamiltonian (\ref{Hamiltonian_1D})
determines the general relation between expectation values of spin and coordinate as
$\left\langle v(t)\right\rangle\equiv d\left\langle x(t)\right\rangle /dt=\alpha
\left\langle \sigma _{z}(t)\right\rangle.$} 

We consider three time-dependent observables:
 $\langle\sigma_{\parallel }\rangle\equiv\langle\sigma_{z}\rangle\cos\theta+
\langle\sigma_{x}\rangle\sin \theta\equiv\left(\langle{\bm\sigma}\rangle\cdot{\bf b}\right),$
$\langle\sigma_{y}\rangle$, and $\langle\sigma_{\perp}\rangle\equiv\langle\sigma_{z}\rangle\sin
\theta -\left\langle \sigma _{x}\right\rangle \cos\theta\equiv\left(\langle{\bm\sigma}\rangle\times{\bf b}\right)_{y}$, 
where the explicit $t$-dependence is omitted 
for brevity. Figure \ref{fig:FIG6}(b) shows the spiral behavior of $\langle\sigma_{y}\rangle$ 
and $\langle\sigma_{\perp}\rangle,$ strongly dependent on the packet width. 
The decreasing with time radius of the spiral here corresponds to the decoherence in the spin subspace.
The final stationary values depend on the width of the packet, that is on the accuracy of
the measurement. For a narrow packet, the displacement of the stationary point from
the initial one is relatively small and increases, same as the maximum radius of the spiral,
with the initial width $w$. 
 
Now we consider asymptotic values of spin components to see the steady states, $t\rightarrow\infty$,
produced by a long measurement. {At long times, $C$ and $S$ defined above Eq.(\ref{g11}) as 
$C\equiv \cos \left( t\sqrt{q^{2}\alpha ^{2}+%
\widetilde{\Delta }^{2}}/2\right)$ and $S\equiv \sin \left( t\sqrt{q^{2}\alpha
^{2}+\widetilde{\Delta }^{2}}/2\right),$ become rapidly oscillating, 
on the scale of the order of $1/t\alpha$, functions of momentum $q$.
As a result, in calculations of integrals containing bilinear forms of these functions one can  
use semiclassical integration rule by substituting in the integrands
$C^2$ and $S^2$ with $1/2$ and $CS$ with zero \cite{LL}. Then, by using} Eqs.(\ref{expHt}) and (\ref{Psi_function}),
with the Green function (\ref{g11}), (\ref{g12}), (\ref{g22}), initial state (\ref{psip}), and 
definition of observables (\ref{sigmat}), we obtain
$\langle\sigma_{\parallel}({\infty})\rangle$ in the form: 
\begin{eqnarray}
&&\langle \sigma_{\parallel }({\infty})\rangle=w
\int_{-\infty}^{\infty}\frac{dq}{2\sqrt{\pi}}e^{-q^{2}w^{2}/4}\times
\label{sigmapar}
\\
&&\frac{\cos \beta \left[ q^{2}\cos \theta +
q\left(\cos^{2}\theta +1\right) +\cos \theta \right] 
+\sin\beta\sin\theta\cos\phi(q\cos\theta  + 1)}
{\left(q+\cos\theta\right)^{2}+\sin^{2}\theta}.\nonumber
\end{eqnarray}

We can study limiting cases of Eq.(\ref{sigmapar}). First limit is the weak spin-orbit coupling, 
that is a broad initial packet, $w\gg \alpha/\widetilde{\Delta}.$ Here 
the range of possible momenta goes to zero, one can substitute $q=0$
in the fraction in the integral to 
obtain $\langle \sigma _{\parallel}({\infty})\rangle=\cos\theta\cos\beta+\sin\theta\cos\phi\sin\beta.$ 
For a strong spin-orbit coupling, that is for a narrow
packet, one can neglect lower than $q^{2}$ powers of $q$ and obtain $%
\left\langle \sigma _{\parallel }\right({\infty})\rangle=\cos \theta \cos
\beta.$ 

The orthogonal component can be written as 
\begin{equation}
\hspace{-1cm}
\langle \sigma _{\perp }({\infty})\rangle=
w\sin \theta
\int_{-\infty}^{\infty} \frac{dq}{2\sqrt{\pi}}e^{-q^{2}w^{2}/4}\hspace{0.1cm}
\frac{\cos \beta (q^{2}+q\cos \theta )+q\cos \phi \sin \beta \sin\theta }
{\left(q+\cos \theta \right) ^{2}+\sin ^{2}\theta }.
\label{sigmaperp}
\end{equation}
In limiting cases considered as below Eq.(\ref{sigmapar}) we obtain 
$\langle \sigma _{\perp }({\infty})\rangle=0$ for weak spin-orbit 
coupling and $\langle \sigma_{\perp}({\infty})\rangle=\sin \theta \cos \beta $ for strong spin-orbit
coupling. The component $\langle \sigma _{y}({\infty})\rangle$ $=0$ for any
coupling strength. 

It is instructive to compare the obtained asymptotic behavior with that 
intuitively expected in a simple case of commuting 
Zeeman and spin-orbit couplings, that is $\theta=0$.  Here the overlap of the 
functions $\Psi_{1}(x|t)=\psi_{0}(x-\alpha t)$ and $\Psi_{2}(x|t)=\psi_{0}(x+\alpha t)$ vanishes on the time scale of
the order of $w/\alpha,$ and the spin state becomes mixed as can be seen in Eq.(\ref{denmat}). 
The expectation values of spin components $\langle\sigma_{z}\rangle=\cos\beta,$
and $\langle\sigma_{y}\rangle=0$ are time-independent. The $x-$component changes from
$\langle\sigma_{x}({0})\rangle=\sin\beta$ to $\langle\sigma_{x}(\infty)\rangle=0$, in agreement
with (\ref{sigmapar}) and  (\ref{sigmaperp}). The spirals in Fig.\ref{fig:FIG6} are transformed
into projection of a log connecting points $(\sin\beta,0,\cos\beta)$ and $(0,0,\cos\beta)$ on the Bloch sphere
and inside the Bloch ball, correspondingly.

%The stationary values $\langle \sigma _{z}({\infty})\rangle=\langle \sigma _{z}(0)\rangle=\cos
%\beta ,$ and $\langle \sigma _{x}({\infty})\rangle=\langle \sigma
%_{y}({\infty})\rangle=0,$ resulting in the above presented 
%$\langle\sigma_{\parallel}({\infty})\rangle$ and $\langle\sigma_{\perp}({\infty})\rangle.$ 
%In the absence of spin-orbit coupling, the spin state
%remains pure with $\langle\sigma_{\parallel}(\infty)\rangle=\langle\sigma_{\parallel}\rangle_{0},$ and 
%$\langle\sigma_{\perp}({\infty})\rangle=\langle\langle\sigma_{\perp}\rangle\rangle_{\infty}$ 
%where $\left\langle \langle ...\rangle \right\rangle_{T}$ corresponds 
%to the double quantum mechanical and time averaging over time interval $T$.

\section{Conclusions and possible extensions}

In summary, we have shown that motion in the spin-orbit coupling fields simulates a 
von Neumann measurement of  spin-1/2 system. 
The spin-orbit dynamics maps spin motion onto a spin-dependent coordinate walk,
thus making distinguishable otherwise hidden spins virtual histories. 
The accuracy of the measurement depends on the resources available in the momentum
space, that is on the spatial width of the initial state. 
We considered two examples of such a procedure: simultaneous measurement of two non-commuting spin
components and measurement of a spin rotating in external magnetic field. 
In the first case, since the virtual Feynman paths can be divided into infinitely
small pieces in the time domain, the produced angular density distribution is time-independent 
at any measurement duration. As a result, any attempt of instantaneous measurement of non-commuting 
spin components fails, and the averages of the spin components 
over the measurement time are observed with the accuracy determined by the width of the
initial wavepacket. In the second case the average of a single spin component corresponding to
spin-orbit coupling axis for a particle moving in one dimension is measured. 
We showed that even if the mass of the particle is infinite,
the initial packet broadens due to spin-orbit coupling, the motion in the coordinated space is complicated,
spin state becomes mixed rather than pure, and decoherence in the spin subspace occurs.

To conclude, we mention several extensions and generalizations of the von Neumann measurement 
procedure seen in the physics of spin-orbit coupling. 

First extension can be related to the manifestation of the Zeno effect \cite{Sokolovski_mes}, 
that is slowing down the dynamics of a constantly measured system. Recent direct calculations indeed demonstrated that  
in the presence of strong driving electric fields \cite{Khomitsky}, or strong
spin-orbit coupling \cite{Nori} the dynamics of the system becomes much slower than 
expected from the linear response approach. It would be of interest to see the relation between
these results and the Zeno effect. 

Second extension can be related to generalizations for other systems and Hamiltonians. 
Recent development in producing synthetic 
spin-orbit coupling allows one to realize the three-dimensional spin-orbit
coupling \cite{Anderson} in the $\left({\bf k}\cdot{\bm\sigma}\right)$ form. Therefore, an
attempt of von Neumann measurement of all three spin components in a single experiment providing
another realization of full qubit monitoring \cite{Ruskov} can be done. Similar problem is
spin-orbit coupling in a system with SU(3) symmetry as can be realized in cold atomic gases \cite{Barnett}. 
Here the linear in momentum Hamiltonian is expressed is terms of eight generators 
of the SU(3) group rather than simply in terms of spin 1 axis projections. As a result, a problem of spin
component measurements becomes more complicated than in the case of spin 1/2, where the number of the SU(2)
group generators is the same as the number of the coordinate axes. 

Another example is provided by holes in two-dimensional semiconductor structures, presenting the realization \cite{Winker03} 
of the $k^3$ rather than linear in $k$ Rashba model. 
Although the von Neumann measurement assumes the linear coupling,
a similar qualitative analysis can be done here.
Taking into account that the spin-orbit splitting here is proportional to
$\gamma k^{3}$, where $\gamma$ is the coupling constant \cite{Winker}, the conditions of accurate measurement
can be reformulated as $\gamma M>w$ and $\gamma T>w^{3},$ and, therefore, a narrow packet (small $w$) is needed for
this purpose.  The realization of this measurement requires a separate discussion. 
 
\section{Acknowledgment}

This work was supported by the University of Basque Country UPV/EHU under
program UFI 11/55, Spanish MEC (FIS2012-36673-C03-01), and ''Grupos
Consolidados UPV/EHU del Gobierno Vasco'' (IT-472-10).


\begin{thebibliography}{99}

\bibitem{Winker03}  Winkler R, \textit{Spin-orbit Coupling Effects in
Two-Dimensional Electron and Hole Systems} Springer Tracts in Modern
Physics (2003)

\bibitem{Zutic04}   Zutic I, Fabian J and Das Sarma S 2004  {\it Rev Mod Phys} \textbf{76} 323

\bibitem{Dyakonov08}  \textit{Spin Physics in Semiconductors} Springer
Series in Solid-State Sciences, Ed by MI Dyakonov, Springer (2008)

\bibitem{Wu10}   Wu M W,  Jiang JH and  Weng MQ 2010 {\it Physics Reports} 
                \textbf{493} 61

\bibitem{GlazovReview} Glazov M M 2012 {\it Physics of the Solid State} \textbf{54} 1

\bibitem{Karimov} Karimov O Z,  John G H,  Harley R T,  Lau W H,  Flatte M E, 
Henini M and Airey R 2003 {\it Phys. Rev. Lett.} \textbf{91} 246601

\bibitem{Rokhinson} See e.g.: Rokhinson L P, Larkina V, Lyanda-Geller Y B, Pfeiffer L N, and West K W
2004 {\it Phys. Rev. Lett.} \textbf{93} 146601 and Schliemann J, Loss D and Westervelt R M 
2005 {\it Phys. Rev. Lett.} 94, 206801 for interesting manifestations of this mutual influence. 

\bibitem{BychkovRashba} Bychkov Y A and Rashba E I 1984 {\it Sov. Phys. JETP Lett.} \textbf{39} 78

\bibitem{Winker} Winkler R 2000 {\it Phys. Rev.} B \textbf{62} 4245

\bibitem{Wang10}   Wang C, Gao C, Jian C-M and Zhai H 2010 {\it Phys. Rev. Lett.} \textbf{105} 160403

\bibitem{Stanescu}  Stanescu T, Anderson B and Galitski V 2008 {\it Phys. Rev.} A \textbf{78} 023616

\bibitem{Lin}   Lin Y-J, Jimenez-Garcia K and Spielman I B 2011 {\it Nature} \textbf{471} 83

\bibitem{Wang}  Wang P, Yu Z-Q, Fu Z, Miao J, Huang L, Chai S, Zhai H and Zhang J 2012 {\it Phys. Rev. Lett.} \textbf{109} 095301

\bibitem{Sommer}  Sommer A, Ku M, Roati G and Zwierlein M W 2011 {\it Nature} \textbf{472} 201

\bibitem{Cheuk}   Cheuk L W,  Sommer A T,  Hadzibabic Z,  Yefsah T, 
                  Bakr W S and  Zwierlein M W 2012 {\it Phys. Rev. Lett.} \textbf{109} 095302

\bibitem{Liu}   Liu X-J, Borunda M F,  Liu X, and Sinova J (2009) 
                {\it Phys. Rev. Lett.} \textbf{102} 046402 

\bibitem{Ozawa}  Ozawa T and Baym G 2012 {\it Phys. Rev.} A \textbf{85} 013612

\bibitem{Xiong}  Liu X-J, Liu Z-X and Cheng M 2013 {\it Phys. Rev. Lett.} \textbf{110} 076401 

\bibitem{ZhaiReview} Zhai H 2012  {\it Int. J. Mod. Phys.} B \textbf{26} 1230001

\bibitem{Tokatly}  Tokatly I V and  Sherman E Ya 2010 {\it Phys. Rev. A} \textbf{87} 041602

\bibitem{Natu} Natu S and Das Sarma S 2013 {\it Phys. Rev. A} \textbf{88} 033613

\bibitem{Wu13} Yu T and Wu M W  2013 {\it arXiv:1309.0727}

\bibitem{Hasan} Hasan M Z and Kane C L 2010 {\it Rev. Mod. Phys.} \textbf{82} 3045

%\bibitem{Tokatly10}   Tokatly I V and  Sherman E Ya 2010 {\it  Annals of Physics} \textbf{325} 1104;
%                      Tokatly I V and  Sherman E Ya 2010 {\it Phys. Rev. B} \textbf{82} 161305

%\bibitem{Koralek09} Koralek  J D, Weber C P,  Orenstein J, 
%Bernevig B A,  Zhang S-C,  Mack S, and  Awschalom D D  2009 {\it Nature} \textbf{458} 610

\bibitem{Gierz} Gierz I, Suzuki T, Frantzeskakis E, Pons S, Ostanin S, Ernst A, Henk J, Grioni M, Kern K and Ast C R
                2009 {\it Phys. Rev. Lett.} \textbf{103} 046803

\bibitem{Yaji} Yaji K, Ohtsubo Y, Hatta S, Okuyama H,
Miyamoto K, Okuda T, Kimura A, Namatame H,
Taniguchi M and Aruga T 2010 {\it Nat. Commun.} \textbf{1} 1016 

\bibitem{Mathias}  Mathias S, Ruffing A, Deicke F, Wiesenmayer M, Sakar I, Bihlmayer G, 
Chulkov E V, Koroteev Y M, Echenique P M, Bauer M and Aeschlimann M 2010
{\it Phys. Rev. Lett.} \textbf{104} 066802

\bibitem{Ibanez} Iba\~{n}ez-Azpiroz J, Eiguren A, Sherman E Y and Bergara A 2012
{\it Phys. Rev. Lett.} \textbf{109} 156401

%\bibitem{RashbaEfros}  Rashba E I and Efros Al L 2003  {\it Phys. Rev. Lett.} \textbf{91} 126405

\bibitem{Allahverdyan13} Allahverdyan A E, Balian R and Nieuwenhuizen T M 2013 {\it Physics Reports} \textbf{525} 1

\bibitem{Flatte}  Berman D H  and  Flatt\'{e} M E 2010  {\it Phys. Rev. Lett.} \textbf{105} 157202

\bibitem{AK}  Arthurs E and Kelly J L, Jr 1965 {\it Bell System Tech} \textbf{44} 725

\bibitem{SHE}   She C Y and  Heffner H, Jr 1966 {\it Phys. Rev.} \textbf{152} 1103

\bibitem{SP0}   Levine R Y and  Tucci R R 1989 {\it Found Phys} \textbf{19} 175

\bibitem{SP9}   Raymer M G 1994 {\it Am J Phys} \textbf{62} 986

\bibitem{SP4}   DAriano G M, Lo Presti P and  Sacchi MF 2002 {\it Phys Lett.} A \textbf{292} 233

\bibitem{SP5}  Allahverdyan  A E, Balian R and  Nieuwenhuizen T M 2010 {\it Physica} E \textbf{42} 339

\bibitem{Sokolovski}  Sokolovski D and Sherman E Ya 2011 {\it Phys. Rev.} A \textbf{84} 030101 (R)

\bibitem{vN}  von Neumann J, \textit{Mathematical Foundation of Quantum Mechanics} (Princeton University Press, Princeton, 1955)

\bibitem{TROT}  Trotter F F 1959 {\it Proceedings of the American Mathematical Society} \textbf{10} 545

\bibitem{Feyn}   Feynman R P and  Hibbs A R, \textit{Quantum Mechanics and Path Integrals} (New York: McGraw-Hill, 1965)

%\bibitem{FOLL} Folland  G B, \textit{Introduction to Partial Differential
%Equations} (Princeton University Press, 1995)

\bibitem{Dugaev}  Similar approach was applied for the Green function in graphene by
Inglot M and Dugaev V K 2011 {\it J. Appl. Phys.} \textbf{109} 123709

\bibitem{ABRAM}   Abramowitz M and Stegun I A \textit{Handbook of Mathematical Functions} 
(New York: Dover Publications 1972).

\bibitem{wire}  Nadj-Perge S, Frolov S M,  Bakkers E\,P\,A\,M and Kouwenhoven L P 2010 {\it Nature (London)} \textbf {468} 1084

\bibitem{plaja} Plaja L and Santos L 2002 {\it Phys. Rev. A} \textbf{65} 035602 

\bibitem{meyrath} Meyrath T P, Schreck F, Hanssen J L, Chuu C S, and Raizen M G
                  2005  {\it Phys. Rev. A} \textbf{71} 041604 

\bibitem{Sokolovski_mes} Sokolovski D 2011 {\it Phys. Rev. A} \textbf{84} 062117; 
                         Sokolovski D 2013 {\it Phys. Rev. D} \textbf{87} 076001 

\bibitem{Shankar} Some useful recepies for Feynman path integrals were given in:  
                  Shankar R {\it Principles of Quantum Mechanics}, (New York: Plenum Press, 2008)

\bibitem{LL} Landau L D and Lifshitz E M {\it Quantum Mechanics Non-Relativistic Theory}, 
            (New York: Butterworth-Heinemann 1981).

\bibitem{Khomitsky}   Khomitsky D V, Gulyaev L V and Sherman E Y 2012
                      {\it Phys. Rev.} B \textbf{85} 125312

\bibitem{Nori} Li R, You J Q, Sun C P and Nori F 2013 {\it Phys. Rev. Lett.} \textbf{111} 086805

\bibitem{Anderson} Anderson B M, Juzeliunas G, Galitski V M and Spielman I B
2012 {\it Phys. Rev. Lett.} \textbf{108} 235301 

\bibitem{Ruskov} Ruskov R, Korotkov A N and M\/{o}lmer K 2012 {\it Phys. Rev. Lett.} \textbf{105} 100506

\bibitem{Barnett} Barnett R, Boyd G R and Galitski V   2012
                  {\it Phys. Rev. Lett.} \textbf{109} 235308


%\bibitem{Levitov03}   Levitov L S and  Rashba E I 2003 {\it Phys. Rev.} B \textbf{67} 115324

%\bibitem{Leurs08}   Leurs B W A,  Nazario Z,  Santiago D I and  Zaanen J 2008
%{\it Annals of Physics} \textbf{323} 907


%\bibitem{Bernevig06}  Bernevig B A,  Orenstein J, and Zhang S-C {\it Phys. Rev. Lett.} \textbf{97} 236601 (2006)

%\bibitem{Slipko}  Slipko V A  and  Pershin Y V 2011 {\it Phys. Rev.} B \textbf{84} 155306

\end{thebibliography}
\end{document}